\documentclass[aps,prl,twocolumn, eqsecnum, groupedaddress]{revtex4}
\usepackage{graphicx}
\begin{document}


\title{A dilute limit of CeAl$_3$: Emergence of the single-ion Kondo scaling }


\author{ C.R. Rotundu, H. Tsujii,}
\altaffiliation[Currently at] { Low Temperature Physics Laboratory, RIKEN, Wako, Saitama 351-0198, Japan }
\author{Y. Takano, and B. Andraka }%
\email{andraka@phys.ufl.edu}
\affiliation{ Department of Physics, University of Florida\\
P.O. Box 118440, Gainesville, Florida  32611-8440, USA}


\date{\today}

\begin{abstract}
Strongly diluted Ce$_x$La$_{1-x}$Al$_3$ alloys have been studied by the low temperature specific heat in order to elucidate the mechanism that determine their ground state. All alloys with Ce concentrations of $0.0005 \leq x \leq 0.1$ show a $S=1/2$ Kondo behavior. However, the single-ion scaling is observed only below $x=0.01$. The true single-ion Kondo temperature is small, 0.2 K, and is identical to that for dilute CeAl$_2$. It is about 20 times smaller than that for CeAl$_3$, indicating that intersite interactions facilitate Kondo-screening in CeAl$_3$ and concentrated Ce$_x$La$_{1-x}$Al$_3$ alloys.
\end{abstract}


\maketitle

\section{}
\narrowtext
CeAl$_3$, the first discovered \cite{Andres} and archetypal heavy fermion compound, remains a subject of extensive experimental and theoretical investigations. For years, this material has been considered as an example of a non-magnetic heavy fermion system with a Fermi-liquid ground state, although short-range magnetic correlations \cite{Barth} have been known to exist below 0.7 K. The results of a more recent alloying study \cite{Andraka} have ignited anew an interest in the true nature of the ground state. Substitution of a few percent of non-magnetic La for Ce results in the appearance of strong anomalies, reminiscent of aniferromagnetic ordering, in thermodynamic properties. In particular, Ce$_{0.8}$La$_{0.2}$Al$_3$ shows a well-pronounced peak in the specific heat at 2.3 K, consistent with ordering of magnetic moments of order 0.3 $\mu$$_B$. However, the elastic neutron scattering \cite{Goremychkin} for the same composition has not detected a long-range order, setting an upper limit for the ordered moment at 0.05 $\mu$$_B$. Instead, a broad inelastic peak was found below 3 K. The neutron scattering results have prompted an interpretation in terms of a single-ion anisotropic Kondo effect, although a subsequent magnetic field investigation \cite{Pietri} has disproved this interpretation. However, the nature of the low temperature state in CeAl$_3$ and (Ce,La)Al$_3$ alloys remains highly controversial and contradictory results are reported. For instance, the zero field $\mu$SR results of Goremychkin et al. \cite{Goremychkin} for Ce$_{0.8}$La$_{0.2}$Al$_3$ seem to be consistent with the anisotropic Kondo model, whereas the $\mu$SR measurements by MacLaughlin et al. \cite{MacLaughlin} for the same nominal composition find sizable frozen moments below the temperature of the peak in the specific heat. 

In order to resolve these inconsistencies and gain further insight into the nature of the ground state of CeAl$_3$ we have previously performed thermodynamic studies of Ce$_x$La$_{1-x}$Al$_3$ over an extended range of concentrations $x$: $0.1 \leq x \leq 1$ \cite{Pietri2}. The overall results were consistent with the Kondo-necklace model \cite{Doniach}. The hexagonal lattice parameter $a$ increases upon the La-substitution causing a weakening of the hybridization between f- and ligand-electrons. The Kondo temperature, $T_K$, estimated from the low temperature entropy, decreases monotonically from approximately 4 K for $x=1$ to about 1 K for $x=0.2$. Moreover, extrapolation of $T_K$ to $x=0$ results in $T_K$ of approximately 0. Such a large variation of $T_K$ due to a relatively small change of one lattice parameter (1.5\%) is quite unexpected considering other extensively studied pseudobinaries involving Ce and La. For instance, across the entire Ce$_x$La$_{1-x}$Pb$_3$ alloy system, $T_K$ is essentially constant \cite{Lin}
 despite some variation of the cubic lattice parameter. Very recent specific heat and magnetic susceptibility data of Ce$_x$La$_{1-x}$CoIn$_5$ were analyzed as a sum of two contributions consisting of the coherent part (due to inter-site correlations) and single-ion Kondo part \cite{Fisk}. The relative weight of these contributions is a function of both temperature and $x$, but $T_K$ is again constant across the series. Moreover, alloys corresponding to $x<0.05$ can be described exclusively by the single-ion  model over all studied temperatures.

For the present study, we have prepared and investigated by specific heat dilute Ce$_x$La$_{1-x}$Al$_3$ alloys, with $x$ ranging from 0.0005 to 0.1. The purpose of this investigation is to answer some of the specific questions raised by previous investigations, among others, whether $T_K$ saturates at a finite value for sufficiently dilute alloys and whether the specific heat can be described by a single-ion isotropic Kondo model in this dilute limit. The anticipated smallness of $T_K$ makes such a study feasible unobscured by the phonon and normal-electron background, but only at temperatures considerably lower than 1 K.

Polycrystalline samples were prepared by arc melting using a multiple step procedure. First, master ingots corresponding to $x=0$ and 0.1 were synthesized. Alloys corresponding to $x<0.1$ were made by successively melting together a more concentrated alloy and $x=0$, such that the ratio of the starting masses of the two alloys was always smaller than 20.  Since the weight losses during the melting and annealing were very small and Al is more volatile than Ce and La, we expect a negligible discrepancy between the nominal and actual concentrations.

f-electron specific heat data presented in Figs 1and 2 were obtained by subtracting the contribution due to normal electrons and phonons, and normalizing the result to a mole of Ce. We have assumed the following temperature dependence of this non-f-electron background: $C{\rm(mJ/Kmol)}=4.6T+0.1213T^3$. The cubic term represents the phonon specific heat for LaAl$_3$, while our linear coefficient is about 7\% smaller than that of the previously reported \cite{Edelstein} Sommerfeld coefficient for LaAl$_3$ (4.95 mJ/K$^2$mol). This small adjustment of the linear coefficient of the background is to avoid negative values for the f-electron specific in the most dilute alloys at temperatures of order 1 K. It has no bearing on the conclusions of this investigation. 
 
\begin{figure}[btp]
\begin{center}
\leavevmode 
\includegraphics[width=0.8\linewidth]{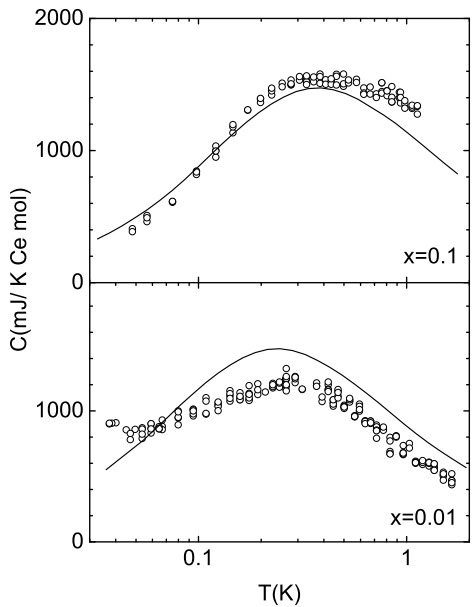} 
\caption{ f-electron specific heat of moderately dilute Ce$_x$La$_{1-x}$Al$_3$ alloys with $x=0.1$ (upper panel) and 0.01 (lower panel). The solid lines are theoretical curves for the $S=1/2$ Kondo model, as described in the text.}
\label{fig1}
\end{center}
\end{figure}
Figure 1 shows the f-electron specific heat for moderately dilute alloys, with nominal Ce-concentrations of $x=0.1$ and 0.01. Both sets of data exhibit broad maxima reminiscent of a Kondo-type specific heat. The maximum occurs at a lower temperature for the $x=0.01$ case. The solid lines are theoretical curves for the $S=1/2$ Kondo model corresponding to $T_K$ of 0.55 and 0.35 K for $x=0.1$ and 0.01, respectively. The experimental points show some systematic deviations from the theoretical curves. In particular, there is an excess specific heat for $x=0.1$ at temperatures larger than 0.5 K, which cannot be eliminated by a reasonable adjustment of the background subtraction. The specific heat values for $x=0.01$ are about 20\% lower than the fit, except for the lowest temperatures. Since another alloy, not shown, obtained from melting together alloys corresponding to $x=0$ and $x=0.01$ also showed lower than expected specific heat values, we believe the actual concentration of Ce is approximately 0.008.

The specific heat data for extremely dilute alloys of $x=0.005$, 0.001, and 0.0005 are shown in Fig. 2. Except for the very low temperature tails, all the experimental data can be described by a single curve, a Kondo fit corresponding to $S=1/2$ and $T_K=0.2$ K. The excess specific heat at the lowest temperatures could already be seen for $x=0.01$ (Fig. 1). If expressed per chemical formula instead of per mole of Ce, the excess specific heat is roughly the same for all alloys. This insensitivity to the Ce concentration, as well as the temperature dependence, strongly suggests that the excess is a Schottky tail of the electric quadrupolar specific heat of
 lanthanum and aluminum nuclei. Such a contribution is expected for atomic nuclei with spins $I$ larger than  $1/2$ in non-cubic environments at miliKelvin temperatures. In indium, for instance, this nuclear contribution \cite{Tang} to specific heat amounts to about 0.4 mJ/K mol at 50 mK. (The electric quadrupol moment, $Q$, of $^{115}$In is 0.83 barns.) This value is comparable to the excess specific heat observed in the present work, taking into account that one mole of dilute Ce$_x$La$_{1-x}$Al$_3$ contains roughly four moles of non $I=1/2$ nuclei: $^{139}$La with $Q$ of 0.22 barns \cite{Carter} and $^{27}$Al with $Q=0.15$ barns \cite{Carter}. We believe, therefore, that the excess specific heat in the low-temperature region of our data is an electric
 quadrupolar contribution of the nuclei, a term totally unrelated to the electrons. In particular, it is
 highly unlikely that it is due to non-Fermi liquid effects.

\begin{figure}[btp]
\begin{center}
\leavevmode 
\includegraphics[width=0.8\linewidth]{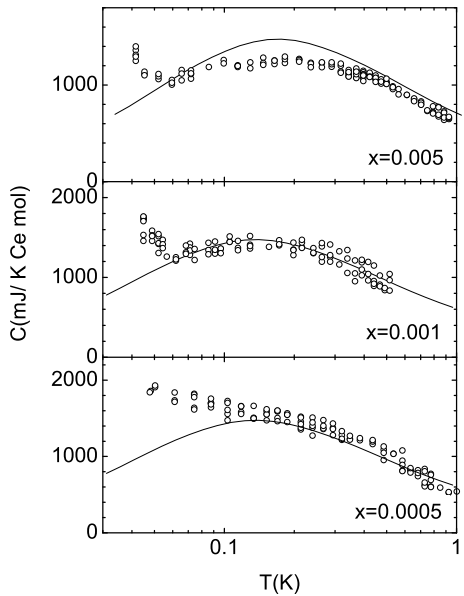} 
\caption{ f-electron specific heat of extremely dilute Ce$_x$La$_{1-x}$Al$_3$ alloys with $x=0.005$ (top), 0.001 (center), and 0.0005 (bottom). The solid lines are theoretical curves for the $S=1/2$ Kondo model, as described in the text.}
\label{fig2}
\end{center}
\end{figure}

The concentration dependence of the Kondo temperature for all alloys studied is shown in Fig. 3. There is a considerable variation of $T_K$ between $x=0.1$ and 0.002, but $T_K$ saturates below this latter concentration. It is difficult to associate these changes of $T_K$ in this relatively dilute range with any changes of  parameters entering the single-ion model. The lattice constants for alloys $x=0.01$ and $x=0.005$ are identical, within the sensitivity of the diffractometer, implying identical hybridization parameters. At the same time, $T_K$ of this latter alloy is almost twice as small as that of the former one. Thus, the observed variation of $T_K$ has to be associated with intersite effects. These intersite effects cease to exist only at extremely low Ce concentrations, of order 0.001, which corresponds to  the average Ce-Ce separation 10 times larger than that in pure CeAl$_3$.  

This result is very surprising in the context of other alloying results reported for Ce-based heavy fermions. For instance, there is a good scaling of specific heat with Ce-content in the aforementioned (Ce,La)Pb$_3$ \cite{Lin}, essentially in the whole alloy parameter space, down to concentrations as low as 0.04. This scaling has been argued to be due to the absence of intersite correlations. Intersite effects are also absent in Ce$_x$La$_{1-x}$CoIn$_5$ \cite{Fisk}, another extensively studied Ce-system, below approximately $x=0.03$. Quite possibly, some clustering in the case of (Ce,La)Al$_3$ is responsible for a much lower value of $x$, in comparison with Ce$_x$La$_{1-x}$CoIn$_5$, below which the intersite effects are not observed.

\begin{figure}[btp]
\begin{center}
\leavevmode 
\includegraphics[width=0.8\linewidth]{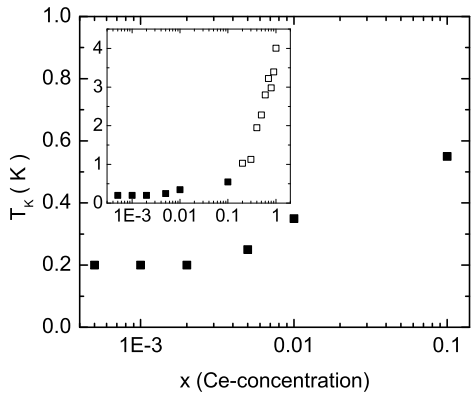} 
\caption{ $T_K$ versus $x$ for Ce$_x$La$_{1-x}$Al$_3$ obtained from the $S=1/2$ Kondo fits. The inset shows $T_K$ for the whole concentration range, open symbols for $x$ between 0.2 and 1 representing the results of ref.\cite{Pietri2}. Note that the method of determining $T_K$ was different, although consistent with the convention used in the present investigation.}
\label{fig3}
\end{center}
\end{figure}

The present study supports the main conclusion derived from our previous investigation \cite{Pietri2} of concentrated Ce$_x$La$_{1-x}$Al$_3$ alloys that the Kondo temperature is a strong function of $x$ and becomes small for dilute alloys. The inset to Fig. 3 presents $T_K$ versus $x$ for the whole alloy parameter space. Note that $T_K$ in the concentrated regime was obtained in a different manner than in the current study. Since the concentrated alloys exhibit magnetic anomalies at low temperatures, $T_K$ was derived by comparing the measured entropy at 3 K with the entropy of the $S=1/2$ Kondo impurity. This method is justified if the entropy removed at temperatures higher than 3 K is primarily via a Kondo effect. A justification to this method was provided by the observation that $T_K$  calculated by using entropies at different temperatures larger than 3 K was leading to similar $T_K$ values. 

We would like to stress that $T_K$ has still considerable variation for $x$ smaller than 0.1. Moreover, $T_K$ increases with the Ce-concentration. Thus, intersite interactions seem to raise $T_K$. It is quite remarkable that the specific heat can be described by the single impurity Kondo model, although $T_K$ itself (the only fitting parameter) is enhanced by the intersite effects. There have been numerous speculations \cite{Millis} on why the Kondo effect survives in concentrated alloys, and collective Kondo screening via RKKY interactions has been proposed. This issue is related to the so-called exhaustion paradox, originally formulated by Nozi\'eres \cite{Nozieres}. At a given temperature $T$, only a fraction $T/T_F$ of conduction electrons is available for the Kondo effect, where $T_F$ is the unnormalized Fermi temperature of order 10$^3$ K. At 1 K, a concentration $x<0.001$ is required for the single-impurity screening to be effective. This critical concentration is roughly consistent with the value of $x$ below which we observe the saturation of $T_K$. However, it is generally believed that increasing the concentration of Kondo centers makes the Kondo screening more difficult, resulting in a decrease of $T_K$. Our experiment shows otherwise.

The Kondo temperature in the dilute regime, 0.2 K, is equal to that of La-diluted \cite{Bader} CeAl$_2$. Is this remarkable result a coincidence, or possibly the true single-impurity $T_K$ is not as sensitive to the Ce-surroundings as generally believed?  The crystal structure, coordination numbers, and the length of bonds are clearly different in the two cases. Ce in the hexagonal structure of LaAl$_3$ has six nearest neighbor Al atoms at a distance of 3.27 \r{A} and six nearest neighbor La atoms at a distance of 4 \r{A}, whereas Ce in the cubic LaAl$_2$ is surrounded by twelve Al atoms at a distance of 3.36 \r{A} and is separated from the four nearest La atoms by 3.50 \r{A}. However, there are other investigations suggesting similar trends in Ce-La alloys when the coherence is gradually suppressed. It has been found that the Kondo temperature in three different compounds, CeAl$_2$, Ce$_3$A$_{11}$, and Ce$_3$Al, is drastically reduced when the size of the sample is reduced to a nanoscale \cite{Chen}. Such a behavior corresponds to a reduction of the free electron path preventing effective collective screening of the magnetic moments.

Our results imply that the true single-ion scaling takes place at concentrations of f-electron ions well below 1\%. Could it then be that the scaling found in (Ce,La)Pb$_3$ alloys does not reflect the true single-ion Kondo behavior? $T_K$ of 3.3 K could be a property of the lattice rather than that of a single ion. Investigations of strongly diluted (Ce,La)Pb$_3$ are under way.

\begin{acknowledgments}
This work has been supported by the U.S. Department of Energy, Grant No. DE-FG02-99ER45748 and by National Science Foundation, DMR-0104240. We thank G.R. Stewart, P. Kumar, and K. Ingersent for stimulating discussions, and K. Ingersent for providing the numerical Kondo fit for $S=1/2$.
\end{acknowledgments}


\end{document}